\documentclass[prd,floatfix,twocolumn,tightenlines]{revtex4}
\usepackage{hyperref}
\usepackage{graphics}
\usepackage{color}
\def\be{\begin{equation}}
\def\ee{\end{equation}}
\def\bea{\begin{eqnarray}}
\def\eea{\end{eqnarray}}

\def\L{{\cal L}}

\def\Lx{{\cal L}_{X}}
\def\Lxzero{{\cal L}_{X0}}

\def\Lxx{{\cal L}_{XX}}

\def\Lp{{\cal L}_{ \phi}}

\def\Lxp{{\cal L}_{X \phi}}
\def\LxP{{{\cal L}_{X}}'}

\def\Hpp{H''}
\def\Hp{H'}
\def\phimp{\frac{\Delta\phi}{\mpl}}

\def\ep{\epsilon}
\def\mpl{M_{pl}}

\def\gaugekappa{\tilde\kappa}
\def\obskappa{\kappa}
\def\gaugeeta{\tilde\eta}
\def\obseta{\eta}
\def\flowb{\alpha}
\def\flowc{\beta}

\begin{document}
\bibliographystyle{prsty}
\title{Reconstructing a general inflationary action }
\author{Rachel Bean $^{1}$}
\author{Daniel J.H. Chung $^{2}$}
\author{Ghazal Geshnizjani $^{2,3}$}
 \affiliation{$^{1}$Department of Astronomy, Cornell University, Ithaca, NY
14853, USA,
\\ $^{2}$Department of Physics, University of Wisconsin,
  Madison, WI 53706, USA,
  \\ $^{3}$Perimeter Institute for Theoretical Physics, Waterloo, Ontario
N2L 2Y5, Canada. }
\date{\today}
\preprint{hep-th/yymmnnn}
\begin{abstract}
If inflation is to be considered in an unbiased way, as possibly
originating from one of a wide range of underlying theories, then observations
need not be simply applied to reconstructing the inflaton potential,
$V(\phi)$, or a specific kinetic term, as in DBI inflation, but rather
to reconstruct the inflationary action in its entirety.  We discuss
the constraints that can be placed on a general single field action from
measurements of the primordial scalar and tensor fluctuation power
spectra and non-Gaussianities. We also present the flow equation
formalism for reconstructing a general inflationary Lagrangian,
$\L(X,\phi)$, with $X=\frac{1}{2}\partial_\mu\phi\partial^\mu\phi$, in
a general gauge, that reduces to canonical and DBI inflation in the
specific gauge $\Lx = c_s^{-1}$.
\end{abstract}
\maketitle
\section{Introduction}
The Cosmic Microwave Background (CMB)
\citep{Page:2006hz,Hinshaw:2006ia,Jarosik:2006ib,Kuo:2006ya} is now
measured with exquisite precision from horizon scales down to a few
arc minutes angular resolution. In combination with large scale
structure surveys \cite{Tegmark:2003uf,Cole:2005sx,Tegmark:2006az},
this allows the primordial spectrum of fluctuations to be
characterized in fine detail
\cite{Spergel:2003cb,Peiris:2003ff,Bridle:2003sa,Spergel:2006hy}.

There has been significant effort to relate the observed primordial
spectrum of fluctuations to the underlying theory that seeded them.
In the context of slow roll inflation
\citep{Guth:1980zm,Linde:1981mu,Albrecht:1982wi}, taking a specific
potential and comparing it to data is a sensible approach to assess
if the theory is consistent (see for example
\citep{Spergel:2003cb,Kinney:2006qm}). An alternative application of
the data, however, is to invert this process, and ÔreconstructÕ what
we can know about the underlying theory
\citep{Copeland:1993ie,Copeland:1993zn,Adams:1994qp,Turner:1995ge,Hoffman:2000ue,
Easther:2002rw,Kinney:2003uw,Peiris:2006ug,Boyanovsky:2007ry,
Powell:2007gu,Lesgourgues:2007gp,Cline:2006db,Ballesteros:2005eg,Cortes:2006ap,
Chung:2003iu,Chung:2005hn,Spalinski:2007ef,Ballesteros:2007te,Destri:2007pv}.

With the introduction of a broader array of inflationary theories, for
example arising out of the Dirac-Born-Infeld action
\citep{Dvali:1998pa} or from k-inflation
\citep{ArmendarizPicon:1999rj, Garriga:1999vw}, the interpretation of
observations has necessarily extended beyond focusing only on the
inflationary potential to including information about the form of the
Lagrangian kinetic term.  Recently, cosmological constraints on brane
inflation models have been studied both in the context of specific
models \citep{Bean:2007hc,Easson:2007dh,Bean:2007eh} and in more
model-independent studies \citep{Peiris:2007gz}.  If inflation is to
be considered without theoretical bias, then the objective must not be
to simply reconstruct the inflaton potential, or a specific kinetic
term, but rather to reconstruct what observations tell us
quantitatively about the effective inflaton action in its entirety. In
this paper, we develop a formalism for such a general inflationary
reconstruction in the context of single field models, and present
explicit analytic techniques for action reconstruction.

In the usual potential reconstruction formalism, the inflationary
observables as a function of scale can be mapped to the behavior of
the inflationary potential $V(\phi)$ as a function of the inflaton
field $\phi$.  If for example the scalar spectral index $n_s(k)$ can
be extracted exactly from data, the shape of the
potential $V(\phi)$ can be deduced in the usual formalism if a reheating scenario is fixed.  However, the reconstruction
of the entire action, including the possibility of non-minimal kinetic
terms, is harder as the action is now a functional of two independent
functions, $X\equiv (\partial \phi)^2/2$ and $\phi$.  To fix this new
functional degree of freedom, in principle a continuous set
of independent data analogous to $n_s(k)$ is needed (i.e. an infinite number of
observables).  Although daunting, the formalism that we present may be
used as a starting point to connect cosmological data to high energy
theories which may have other phenomenological, theoretical, and
aesthetic constraints.

The analytic form of the non-minimal kinetic actions consistent with
data can be written in a surprisingly simple form given in section
\ref{srrecon} by Eq.~(\ref{generalaction}).  Each consistent action is
simply a manifold parameterized by $X$ and $\phi$ satisfying certain
derivative conditions on a one dimensional submanifold which
represents the data.  Furthermore, using the Hamilton-Jacobi
formalism, we extend the inflationary flow parameter approach to
describe the evolutionary trajectories of general actions. This
involves introducing three hierarchies of flow parameters to describe
the evolution of a general action without using the specific
restriction of field redefinition used in canonical and DBI
inflation. These equations hold for all single field inflationary
scenarios, whether or not slow roll conditions are met.

The importance of including kinetic terms in the inflationary
reconstruction program cannot be overemphasized in light of recent
theoretical and expected experimental advances.  Inflationary models
with non-minimal kinetic terms are able to produce large non-Gaussian
behavior for the curvature perturbations without ruining other
inflationary observables \citep{Alishahiha:2004eh} and predictions of
non-Gaussian signatures for specific models have been established, for
example in DBI inflation
\cite{Creminelli:2004yq,Seery:2005wm,Chen:2006nt,Maldacena:2002vr,Bean:2007eh}.
Indeed, the search for such non-Gaussian effects is one of the primary
current activities in observational cosmology, for example
\citep{Komatsu:2003fd,Spergel:2006hy,Yadav:2007yy}.  Non-Gaussianity
detections open up the possibility of establishing which non-minimal
kinetic terms may exist for inflationary models. The formalism that we
present here will be useful for this purpose.

The explanation of any future (or current) observations of
non-Gaussianities can also be checked in the context of single
field inflation through the attendant modification of the tensor
spectral index consistency relationship \cite{Garriga:1999vw}.  The latter can be deduced
experimentally from observations of tensor perturbations implied by
CMB B-mode polarization measurements.  However, one advantage of
emphasizing the non-Gaussianity connection with non-minimal kinetic
terms is that the possibility of a large non-Gaussian contribution is
generically independent of the single field paradigm.  On the other hand,
the tensor spectral index consistency relationship changes for
multifield inflationary models.

The order of our presentation will be as follows.  In section
\ref{kinetic}, we review and clarify the physics of how non-minimal
kinetic terms contribute to non-Gaussianities, whose possible future
observation is one of the strongest motivations for developing the
action reconstruction formalism.  In \ref{backg} we outline the
general equations for the background evolution. We discuss the
conditions for slow roll inflation in a general action in section \ref{sr} .  
In \ref{pert} we summarize how the generalized flow 
parameters relate to the properties of the primordial power spectrum
and discuss how properties of a general action can be distinguished
from canonical inflation using cosmological observations. In section
\ref{srrecon} we establish how the general action can be
reconstructed from measurements of the lowest order flow parameters
in the slow roll regime, and in \ref{flow} we extend the
inflationary flow parameters \citep{Liddle:2003py} to describe a
general inflationary action. In \ref{conc} we summarize our findings
and discuss their implications.

Throughout this paper with the exception of section \ref{srrecon}, we
use the usual reduced Planck scale conventions of $\mpl^2 = (8\pi
G)^{-1}\approx (2.4\times 10^{18} \mbox{GeV})^2$.  In section VI, we
will use geometricized units and set $M_{pl}=1$ for simplicity in
notation.

\section{Non-Gaussianity and Non-minimal Kinetic Terms}\label{kinetic}
Although there have been many previous works
\citep{Copeland:1993ie,Copeland:1993zn,Adams:1994qp,Turner:1995ge,Hoffman:2000ue,Easther:2002rw,Kinney:2003uw,Peiris:2006ug,Boyanovsky:2007ry,Powell:2007gu}
on inflationary potential reconstruction, there are relatively
fewer works on trying to reconstruct kinetic terms
\citep{Peiris:2007gz}. As explained in the introduction, one of the
main motivations for focusing on non-minimal kinetic terms is its
importance to non-Gaussian observables, whose search is an active area
of research in observational cosmology.  In this section, we explain
how non-minimal kinetic terms can generate observable
non-Gaussian statistics. Most of this section is devoted to
summarizing and clarifying the literature which is particularly
relevant for this paper.

All field correlation functions of a \emph{non-interacting} field
theory can be reduced to the information in the two-point correlation
function similar to the statistics of a classical Gaussian random
variable. During slow roll inflation, the energy density fluctuations
of the inflaton $\phi$ (the dominant energy component) are approximately
\begin{equation}
\delta\rho_{\phi}(x)\sim V'(\phi_{0})\delta\phi(x)\label{eq:linearrelationship}\end{equation}
where $V(\phi)$ is the inflaton potential, $\phi_{0}$ is the classical
time dependent background homogeneous inflaton field, and $\delta\phi(x)$
is the quantum fluctuating inflaton field degree of freedom. Hence,
if $\delta\phi$ fluctuations (which eventually decohere to become
classical) can be described by a non-interacting field theory, then
the statistics of $\delta\rho_{\phi}$ will also will be Gaussian
since by the linear relationship given in Eq.~(\ref{eq:linearrelationship}),
it inherits the statistics of $\delta\phi$.

In slow roll inflationary theories with minimal canonical kinetic
terms, the inflaton field still interacts non-trivially with gravity,
leading to non-Gaussian statistics of $\delta\rho_{\phi}$. However,
because the energy density fluctuations are small, the
gravity-mediated self-interactions are typically small. Furthermore,
the slow roll constraints also phenomenologically forces the coupling
constants in the self-interaction terms of the potential to be small,
suppressing non-gravity-mediated self-interactions. One typical
characterization of the non-Gaussian statistics is the 3-point
function\bea
&&\langle\zeta(\tau,\vec{k}_{1})\zeta(\tau,\vec{k}_{2})\zeta(\tau,\vec{k}_{3})\rangle = (2\pi)^{7}\delta^{(3)}(\vec{k}_{1}+\vec{k}_{2}+\vec{k}_{3}) \times \nonumber
\\
&& \ \ \ \left[ P^{\zeta}(k_{1}+k_{2}+k_{3}) \right]^{2}\frac{\mathcal{A}(\vec{k}_{1},\vec{k}_{2},\vec{k}_{3})}{\prod_{i}k_{i}^{3}}\label{eq:ampdef} \ \ \ \ \eea
where $\mathcal{A}$ is a smooth function with dimension $[k]^{3}$,
$\zeta$ is the scalar perturbation which in the $\delta\phi=0$ gauge
parameterizes the spatial metric as $\exp(2\zeta)|d\vec{x}|^{2}$,
$\tau$ is conformal time when all the scales are far outside of the
horizon during inflation, and $P_{k}^{\zeta}$ is the two-point
function power spectrum.\footnote{Here, $P^{\zeta}(k)$ is normalized
such that in the usual slow roll models, it reduces to $P^{\zeta}(k)=
V/(24\pi^2 M_{pl}^4 \epsilon)$ where $\epsilon\equiv
M_{pl}^2(V'(\phi)/V(\phi))^2/2$.}  To linear order, the scalar
perturbation reduces to the linearly gauge invariant function
$\zeta=-\Psi-\frac{H}{\dot{\phi}_{0}}\delta\phi$, where $\Psi$ is
scalar perturbation appearing in the line element ($dt^{2}(1+2\Psi)$)
and $H$ is the expansion rate. The literature often characterizes the
amplitude $\mathcal{A}$ at either the
$\vec{k}_{1}=\vec{k}_{2}=\vec{k}_{3}$ limit (equilateral triangle), or
$|\vec{k}_{1}|\ll|\vec{k}_{2}|,|\vec{k}_{3}|$ limit (squeezed or local
limit). In each of these cases, a dimensionless quantity $f_{NL}$ can
be defined by the relation \citep{Chen:2006nt}
\begin{equation}
\mathcal{A}(\vec{k}_{1},\vec{k}_{2},\vec{k}_{3})\equiv-\frac{3}{10}f_{NL}^{\mbox{equil or local}}\sum_{i}k_{i}^{3}\label{eq:fnldef}\end{equation}
where the definition is motivated by the characterization of non-Gaussianities
by a non-general ansatz $\zeta=\zeta_{G}-\frac{3}{5}f_{NL}^{\mbox{equil or local}}(\zeta_{G}^{2}-\langle\zeta_{G}^{2}\rangle)$
which is valid only when the non-Gaussian variable $\zeta$ is related to
the Gaussian variable $\zeta_{G}$ by a local field redefinition.

Having a $f_{NL}^{\mbox{equil or local}}>0$ in the sign convention of
Eq.~(\ref{eq:fnldef}) corresponds to having more hot spots in the CMB
anisotropies compared to the case with $f_{NL}=0$. To see this, note
that the observed anisotropies on large scales is approximately
$\frac{\Delta T}{T}\approx-\frac{1}{5}\zeta$ due to Sachs-Wolfe
effect.  Hence, the temperature one point function $P(\Delta T/T)$ should
behave approximately as
\begin{equation}
 \ln P \propto \left[\frac{\Delta
T}{T}-3 f_{NL} \left(\left(\frac{\Delta T}{T} \right)^{2}-\langle \left(\frac{\Delta
T}{T} \right)^{2}\rangle\right)\right]^{2},
\end{equation}
which makes the probability of having $\Delta T/T$ larger than the
standard deviation a bit higher. Note that the sign convention of
\cite{Chen:2006nt,Maldacena:2002vr} is opposite to the sign convention
of \citep{Yadav:2007yy}.  Furthermore, the non-zero value of $f_{NL}$
measured by \citep{Yadav:2007yy} is in the squeezed limit of $k_{1}\ll
k_{2},k_{3}$ which is less sensitive to the non-minimal kinetic term
as pointed out by several papers (e.g. see
\citep{Creminelli:2004yq,Seery:2005wm,Chen:2006nt,Maldacena:2002vr}).
To leading order, $f_{NL}$ can be related to the scalar
perturbation spectral index as
\begin{equation}
f_{NL}^{\mbox{local}}\sim(n_{s}-1)
\end{equation} which is suppressed (as
these are proportional to the slow roll parameters) with a negative
sign in the current sign convention if $n_{s}<1$. Hence, it is
interesting that the result of \citep{Yadav:2007yy} is \emph{not}
likely to be explained by something like DBI inflation or more
generally, by non-minimal kinetic term effect only. For slow roll
inflationary models, this small $f_{NL}$ proportional to the slow roll
parameter is generic \citep{Maldacena:2002vr}.

One idea to make $f_{NL}$ large from non-gravitational
self-interactions that people did not pay much attention to before
\citep{Alishahiha:2004eh} was that generically self-interactions can
be made large without preventing inflation if the self-interactions
come from non-minimal kinetic terms. From an intuitive point of
view, one sees that if the inflaton Lagrangian has the
form\begin{equation}
\mathcal{L}_{\mbox{intuition}}=f((\partial\phi)^{2},\phi)(\partial\phi)^{2}-m^{2}\phi^{2}\label{eq:intuitievelagrangian}\end{equation}
where $m$ is a mass parameter and $f(a,b)$ is a function which has a
large numerical value, say $Z\gg1$, along a particular classical
solution, then by redefining the field to be
$\tilde{\phi}\equiv\sqrt{Z}\phi$, we have numerically\begin{equation}
\mathcal{L}_{\mbox{intuition}}\sim(\partial\tilde{\phi})^{2}-\frac{m^{2}}{Z}\tilde{\phi}^{2}.\end{equation}
This makes the effective potential even flatter than the situation in
which $Z$ was of order $1$, which in turn helps in meeting the
phenomenological inflationary conditions. At the same time, if
$Z=f((\partial\phi)^{2},\phi)$ is large, then there are
non-renormalizable self-interactions of $\phi$ in
Eq.~(\ref{eq:intuitievelagrangian}) that are large, and hence the
expected non-Gaussianities can be large without spoiling inflation.

Therefore, one key to obtaining large non-Gaussianities in inflationary
models with a single scalar field is to consider modifications of
the kinetic term. To see how $f_{NL}$ can be related to the nonrenormalizable
interactions appearing in the kinetic sector, consider a dimension
8 non-renormalizable interaction of the form\begin{equation}
S_{int}=\int d^{4}x\sqrt{g}\frac{c}{\Lambda^{4}}(\partial_{\mu}\phi\partial^{\mu}\phi)^{2}.\label{eq:dimension8}\end{equation}
 After expanding $\phi$ as $\phi_{0}(t)+\delta\phi(x)$, we see that
Eq.~(\ref{eq:dimension8}) contains a cubic self-interaction \begin{equation}
S_{int}\ni\int d^{4}xa^{3}\frac{4c}{\Lambda^{4}}\dot{\phi}_{0}(t)\delta\dot{\phi}^{3}(x)\label{eq:interactionlagrangian}\end{equation}
where the dot denotes the partial derivative with respect to a comoving
observer's proper time and $\phi_{0}(t)$ is governed by the quadratic
Lagrangian\begin{equation}
S_{2}=\int d^{4}x\sqrt{g}\left[\frac{1}{2}(\partial_{\mu}\phi\partial^{\mu}\phi)-V(\phi)\right].\end{equation}
For the present discussion, we will assume that $V(\phi)$ energy
density and pressure are dominant during inflation. It is interesting
to note that the cubic derivative interaction of Eq.~(\ref{eq:interactionlagrangian})
is induced by having \emph{both} a dimension 8 short distance operator
(Eq.~(\ref{eq:dimension8})) and a nonvanishing time dependent background
field $\phi_{0}(t)$. The second condition is required because local
Lorentz invariance forces Eq.~(\ref{eq:dimension8}) to have a $Z_{2}$
symmetry which needs to be spontaneously broken by $\phi_{0}(t)$
to obtain a cubic interaction.

If the scalar metric fluctuation can be neglected, then the dominant
contribution to the 3-point function would simply come from
Eq.~(\ref{eq:interactionlagrangian}).  However, as is well known by
now (\cite{Seery:2005wm,Chen:2006nt}), the scalar metric perturbations
induce significant contributions to $\langle\zeta\zeta\zeta\rangle$
proportional to $c$ in Eq.~(\ref{eq:dimension8}).  To account for the
scalar metric perturbations, it is often more convenient to choose the
foliation of spacetime with spacelike 3-surfaces in which
$\delta\phi=0$ and use the ADM formalism to construct the interacting
Lagrangian and Hamiltonian \citep{Maldacena:2002vr} in terms of the
metric component $\exp(2\zeta)$ characterizing 3-metric as
$\exp(2\zeta)d|\vec{x}|^{2}$. Explicitly, the leading interaction
Lagrangian can be written as \cite{Seery:2005wm}
\begin{equation} \mathcal{L}_{I}\sim
a^{3}\epsilon
u\left[\frac{-2}{3}\frac{\dot{\zeta}^{3}}{H}+8a^{2}\dot{\zeta}^{2}\partial^{-2}\dot{\zeta}\right]
\end{equation}
where $u=\frac{-8c\dot{\phi}_{0}^{2}}{\Lambda^{4}}$ for the case of
Eq.~(\ref{eq:dimension8}). The second term (whose peculiar non-local
form comes from solving the non-local constraint equations of gravity)
turns out to dominate in contribution to
$\langle\zeta\zeta\zeta\rangle$ over the local interactions
represented in the first term in the limit that the slow roll
parameters vanish.  This indicates that the metric perturbations
cannot be neglected in computing the 3-point functions for non-minimal
kinetic terms. What is intriguing about this is that non-Gaussianities
may in fact be a sensitive probe of gravity.  Although we will leave
investigations of this issue to a future work, it is interesting to
note that the modifications of gravity proposed by
\cite{Afshordi:2006ad,Afshordi:2007yx} directly changes the
gravitational constraint equations which the scalar metric
perturbations are sensitive to.
\begin{figure*}
\begin{centering}
\includegraphics{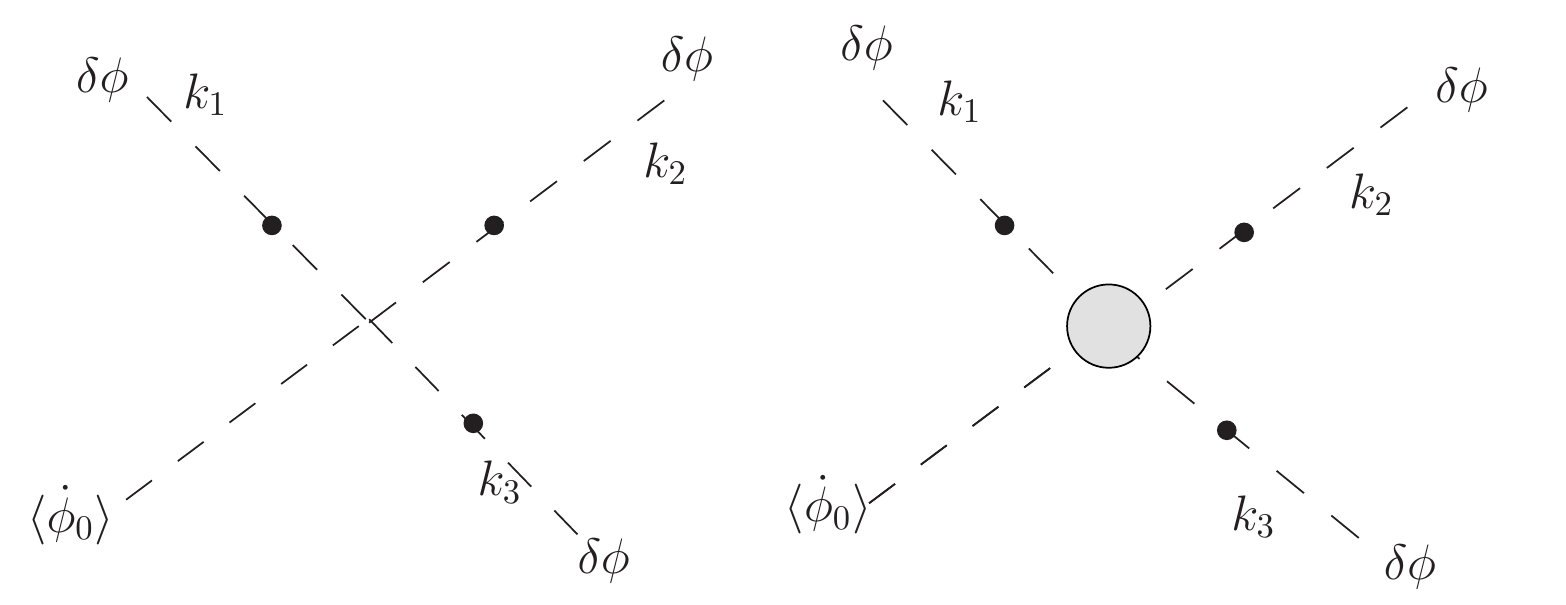}\par
\end{centering}
\caption{\label{fig:Dimension-8-operator}Dimension 8 kinetic operator interaction
of $\delta\phi$ contribution to the 3-point function of $\langle\zeta(t,\vec{k}_{1})\zeta(t,\vec{k}_{2})\zeta(t,\vec{k}_{3})\rangle$.The small dots indicates the fact that the $\delta\phi$ propagator is a dS propagator (i.e. there is an interaction with the background
classical homogeneous gravitational field leading to a time dependent
mass). The blob on the right indicates that it is an interaction term
(partly non-local) arising from the presence scalar metric fluctutations.}
\end{figure*}

The 3-point function is computed perturbatively as
\bea \langle\mathcal{O}(t)\rangle & = & \langle
e^{i\int_{t_{0}}^{t}dt'H_{I}(t')}\mathcal{O}_{int}(t)e^{-i\int_{t_{0}}^{t}dt'H_{I}(t')} \!\rangle\\
& \approx &
\langle\mathcal{O}_{int}\rangle+i\int_{t_{0}}^{t}dt'\langle[H_{I}(t'),\mathcal{O}_{int}(t)]\rangle\eea
where
$\mathcal{O}=\zeta(t,\vec{k}_{1})\zeta(t,\vec{k}_{2})\zeta(t,\vec{k}_{3})$,
$H_{I}$ is the interacting Hamiltonian $(H_{I}=-\int
d^{3}x\mathcal{L}_{I}$), and the first term is vanishing for our
observable. Hence, noting that in the spatially flat gauge,
$\zeta\sim\frac{-H}{\dot{\phi}_{0}}\delta\phi$, and near the slow roll
limit
$\dot{\phi}_{0}\sim\sqrt{2\epsilon_V}HM_{p}\mbox{sgn}(\dot{\phi}_{0})$
(where $\epsilon_V \equiv M_{pl}^2 (V'(\phi)/V(\phi))^2/2$),
the leading effect of the non-minimal kinetic term on the
3-point function can be represented by the diagram in
Fig. \ref{fig:Dimension-8-operator}.  Because the background spacetime
has spatial translational invariance, there will still be an overall
3-momentum conservation
$(2\pi)^{3}\delta^{(3)}(\vec{k}_{1}+\vec{k}_{2}+\vec{k}_{3})$ in the
computation. However, because of the propagator being in dS spacetime
patch which is not time translationally invariant, the time integral
will not conserve $\sum_{i}|\vec{k}_{i}|$ but will instead lead to a
factor of $1/H$. Since the vertex of the first diagram in
Fig. \ref{fig:Dimension-8-operator} can be read off from
Eq.~(\ref{eq:interactionlagrangian}) as
$\frac{c\dot{\phi}_{0}}{\Lambda^{4}}$, we can estimate
\begin{eqnarray}
\langle\zeta(t,\vec{k}_{1})\zeta(t,\vec{k}_{2})\zeta(t,\vec{k}_{3})\rangle
\sim
(2\pi)^{3}\delta^{(3)}(\vec{k}_{1}+\vec{k}_{2}+\vec{k}_{3})\times \ \
\nonumber \\
\frac{1}{H}\times
\frac{c\dot{\phi}_{0}}{\Lambda^{4}}\times(\frac{H}{\dot{\phi}_{0}})^{3}
\times(H^{2})^{3}h(\vec{k}_{1},\vec{k}_{2},\vec{k}_{3}) \ \ \
\\  =
(2\pi)^{7}\delta^{(3)}(\sum_i\vec{k}_{i})\frac{c\dot{\phi}_{0}^{2}}{\Lambda^{4}}\left[
  \left(\frac{H}{\dot{\phi}_{0}}\frac{H}{2\pi}\right)^{2}\right]^{2}h(\vec{k}_{1},\vec{k}_{2},\vec{k}_{3})\label{eq:secondlineform}\ \
\end{eqnarray}
where $h(\vec{k}_{1},\vec{k}_{2},\vec{k}_{3})$ is a Fourier transform
related kinematic function which scales as $1/k^{6}$ in the
equilateral triangle limit of
$|\vec{k}_{1}|=|\vec{k}_{2}|=|\vec{k}_{3}|=k$, and the other factors in
the first equation can be explained as follows. The factor of $1/H$ comes
from the integral $\int dt$ (the effective interaction having a time
scale of order $1/H$, and this is what results from the
nonconservation of $\sum_{i}|\vec{k}_{i}|$), the factor of
$c\dot{\phi}_{0}/\Lambda^{4}$ comes from the interaction vertex,
$(H/\dot{\phi}_0)^3$ comes from the relationship
$\zeta\sim\frac{-H}{\dot{\phi}_{0}}\delta\phi$, and $(H^{2})^{3}$
comes from each of the external $\langle \delta \phi \delta \phi
\rangle$ propagators in Fig. \ref{fig:Dimension-8-operator} being
proportional to $H^{2}$ (the well known massless dS propagator
scaling). The second line follows from trying to express the result as
Eq.~(\ref{eq:ampdef}), where
\begin{equation}
P_{k}^{\zeta}\approx\left(\frac{H}{\dot{\phi}_{0}}\frac{H}{2\pi}\right)^{2}\end{equation}
 to leading slow roll and non-minimal kinetic term vertex order.

Comparing Eq.~(\ref{eq:secondlineform}) with Eq.~(\ref{eq:fnldef}) and
Eq.~(\ref{eq:ampdef}), we find for equilateral momentum triangle
configuration \begin{equation} f_{NL}^{\mbox{equil}}\sim
c\frac{\dot{\phi}_{0}^{2}}{\Lambda^{4}},\label{eq:equilfinalestimate}\end{equation}
up to a minus sign which we cannot predict with the current detail of
estimation, which leaves out the gravitational effects (e.g. the
second diagram of Fig.~\ref{fig:Dimension-8-operator}).\footnote{For a
systematic diagrammatic approach to computing correlation functions in
the $\delta N$ formalism, see \cite{Byrnes:2007tm}.} The inclusion of
the gravitational effects \cite{Seery:2005wm,Chen:2006nt}
yields\begin{equation} f_{NL}^{\mbox{equil}} \approx
0.28\frac{2X\mathcal{L}_{XX}}{\mathcal{L}_{X}}\end{equation} which for
Eq.~(\ref{eq:dimension8}) yields \begin{equation}
f_{NL}^{\mbox{equil}}\approx
0.28\frac{8c\dot{\phi}_{0}^{2}}{\Lambda^{4}}\end{equation} in
agreement with Eq.~(\ref{eq:equilfinalestimate}).

Before closing this section, we would like to also comment that Eq.
(\ref{eq:equilfinalestimate}) can be rewritten in terms of the
potential slow roll parameters as \begin{equation} f_{NL}^{\mbox{equil}}\sim
c\epsilon_{V}\frac{H^{2}}{\Lambda^{2}}\frac{M_{pl}^{2}}{\Lambda^{2}}.
\label{eq:leadingperturbative}\end{equation}
This expression is interesting because although the dimension 8
operators of the form of Eq.~(\ref{eq:dimension8}) are generically
expected to exist in conventional effective field theories with
$c\sim\mathcal{O}(1)$ because $\Lambda$ then is the cutoff scale, the
validity of the effective field theory description requires
\begin{equation} H^{2}M_{pl}^{2}<\Lambda^{4}.\end{equation}
Eq.~(\ref{eq:leadingperturbative}) would then imply that the $f_{NL}$
contribution from perturbatively treated non-minimal kinetic operators
would be suppressed by $\epsilon_{V}$ in a typical effective field
theory. However, there are apparently situations such as in DBI
inflationary models in which
$\frac{2X\mathcal{L}_{XX}}{\mathcal{L}_{X}}$ can be large yet a
sensible effective field theory description exists
\cite{Chen:2005fe,Silverstein:2003hf}.  Such scenarios would still
give a large value for $f_{NL}^{\mbox{equil}}$ due to non-minimal
kinetic term interactions.

\section{Background evolution}\label{backg}
Consider a general action of a single scalar field with a Lagrangian of the
form $\L(X,\phi)$ where $X={1\over
2}\partial_\mu\phi\partial^\mu\phi$ is the canonical kinetic term. One can describe
the $\phi$ field by a hydrodynamical fluid  in the following way:
\begin{equation}
T_{\mu\nu}=(p+\rho) u_\mu u_\nu -pg_{\mu\nu},
\end{equation}
where
\bea
p(X,\phi) &\equiv& \L(X,\phi),
\\
\rho(X,\phi) &\equiv&2X\Lx-\L(X,\phi), \label{density}\\
u_\mu&\equiv& {\partial_\mu \phi \over \sqrt{2X}},
\eea
 where $\Lx \equiv \partial \L /\partial X$. In the homogenous limit that $X={1\over 2} \dot{\phi}^2$ Eq.~(\ref{density}) simplifies to
\begin{equation} \label{gdensity}
\rho(X,\phi)= \dot{\phi}{\partial\L\over\partial \dot{\phi}}-\L.
\end{equation}

In this paper we assume that null energy condition $\rho+p>0$, is
satisfied, such that
\begin{equation}
2X{\partial \L\over \partial X}>0 \label{pxeq}.
\end{equation}

The Friedmann, acceleration and continuity equations for the
background are
\bea
H^2&=&{1\over 3\mpl^2}\left(2X\Lx-\L\right), \label{fr1}
\\
{\ddot{a}\over a}&=&-{1\over 3\mpl^2}(X\Lx+\L),\label{fr2}
\\
\dot{\rho}&=&-3H(\rho+p),\label{cont}
\eea
where $a(t)$ is  the scale factor, $H$ is Hubble's
constant$\equiv \dot{a}/a$.
Accelerative expansion  requires
\bea\label{accLx}
0<\frac{X\Lx}{-\L}<1,
\eea
with $\L<0$.
The resulting equation of motion for the scalar field is
\bea
\dot{X}&=&\frac{\sqrt{2X}c_s^2}{\Lx}\left(\Lp-2X\Lxp-3H\sqrt{2X}\Lx\right), \label{eom} \ \ \
\eea
where throughout we choose the sign of $\sqrt{2X}$ to be same as $\dot{\phi}$. $c_s^2$ is defined as
\bea\label{defcs}
c_s^2\equiv{p_{X}\over \rho_X}=\left(
1+2\frac{X \Lxx}{\Lx}\right)^{-1}.
\eea
As we will see in the next section, it turns out to be the adiabatic sound speed for inhomogeneities.  Requiring $c_s^2\leq 1$ and the positivity of $\Lx$, giving $c_s^2>0$ from Eq.~(\ref{pxeq}) yields,
\begin{equation}
\Lxx>0.
\end{equation}
Combining Eq.~(\ref{fr1}) and Eq.~(\ref{cont}) we can write the  kinetic variable as a function  $H$ and $\Lx$,
\bea
\sqrt{2X} &=& -\frac{2\mpl^2}{\Lx}\Hp\label{Hprime},
\eea
where a prime denotes a total derivative with respect to $\phi$,
\bea
\Hp \equiv \frac{dH}{d\phi}=\frac{1}{\sqrt{2X}}\frac{dH}{dt},
\eea
where
\bea
\frac{d}{dt}=\dot{X}\frac{\partial}{\partial X}+ \sqrt{2X} \frac{\partial}{\partial\phi}. \label{partial}
\eea
Using Eq.~(\ref{eom}), we can therefore describe all time derivatives in terms of partial derivatives with respect to $\phi$ and $X$.

Notice that if  $\L(X,\phi)$ is known, $\Lx(X,\phi)$ can be derived
and inserted in Eq.~(\ref{Hprime}) to solve for $X(\Hp, \phi)$,
which then can be substituted back in Eq.~(\ref{fr1}) to obtain a
nonlinear first order differential equation for $H(\phi)$ which
similar to canonical actions corresponds to the Hamilton-Jacobi (HJ)
equation for the general action \cite{Salopek:1990jq} \begin{equation}
3\mpl^2H^2(\phi) =
\frac{4\mpl^{4}\Hp^2}{\Lx\bigl(X(\Hp,\phi),\phi\bigr)}
-\L\bigl(X(\Hp,\phi),\phi\bigr) \label{HJ}.\ \ \ \ \ \ \ \end{equation}
We can rewrite the HJ equation in terms of a new parameter $\epsilon$,
\bea
3\mpl^2H^2\left(1-\frac{2\ep}{3}\right)=- \L,
\eea
where
\bea
\epsilon &\equiv& {3(\rho+p)\over 2 \rho}=-\frac{\dot{H}}{H^{2}} \label{epsilonH}
\eea
and it can also be written in following formats in terms of parameters in the action,  
\bea
\epsilon=\frac{3}{2-\frac{\L}{X \Lx}}=\frac{2\mpl^2}{\Lx} \left(\frac{\Hp}{H}\right)^2
.\label{epsilon}
\eea
The physical relevance of $\ep$ is more clearly seen given
\bea
\frac{\ddot{a}}{a}=(1-\epsilon)H^2\label{accel},
\eea
which implies that the accelerative expansion condition Eq.~(\ref{accLx}) can also be written for $H$ as  $0<\ep< 1$.

To design a successful inflationary scenario, it is necessary to first
address the flatness and horizon problem, for which it suffices to
have \citep{Liddle:1994dx}
 \begin{equation} \label{ahcondition}
 \tilde{N}\equiv \ln\left|{a(t_{en})H(t_{en})\over a(t_{in})H(t_{in})}\right|>
 \ln\left((1+z_{eq})^{-1/2}{T_{re}\over T_0}\right),
 \end{equation}
 where $T_0$, $T_{re}$ and $z_{eq}$ are the CMB temperature today, the
 reheating temperature, and the redshift at time of matter-radiation
 equality, respectively. The left hand side describes the logarithmic
 shrinkage of Hubble radius in the comoving frame. To be consistent
 with observations, the reheating temperature has to be above
 nucleosynthesis scales which yields $ \tilde{N}\geq 24$, but if one
 assumes that a reheating temperature is as high as the GUT scale then
 a larger lower limit, $ \tilde{N}\geq 80$, is obtained.

A commonly used measure of inflation is the number of e-folds of
inflation, $N_e$, defined as
\bea
N_e &\equiv& \ln{a(t_{en})\over a(t_{in})}=-\int_{t_{en}}^{t_{in}}Hdt \nonumber \\
&= &\frac{1}{\mpl} \int_{\phi_{in}}^{\phi_{en}}\sqrt{\frac{\Lx}{2\ep}}d\phi ,
\eea
in which $t_{in}$ and $t_{en}$ are the start and end of inflation, and we
choose $N_e$ to  increase as one goes backwards in time from the end
of inflation i.e  $dN_e= -Hdt$.

  If $H$ is changing slowly during inflation, then $N_e \sim  \tilde{N}$ and the constraint on the Hubble radius shrinkage can be satisfied simply by requiring $N_e >  \ln\left((1+z_{eq})^{-1/2}{T_{re}\over T_0}\right)$.  In general, however, once one enforces the null energy condition $\rho+p>0$, since $\dot{H}$
is negative, inevitably $N_e$ is $\tilde{N}+\ln{H_{in}\over
H_{en}}$, and hence is larger than $\tilde{N}$: \bea N_e>{\tilde{N}\over
1-\ep_{min}}, \eea which implies that for scenarios in which $\ep$ is not
close to zero one must obtain significantly larger number for
e-folding to solve the horizon problem.
 Keep in mind that imposing a higher reheating temperature, and requiring the initial condition
$\rho_{in}<M_{pl}^4$ at the same time, will only restrict $\ln[H(t_{in})/ H(t_{en})]< 38$, and only marginally constrain $\ep$
to be about $10\%$ less than one.
Regardless of the null energy condition,
meeting the observable constraint Eq.~(\ref{ahcondition}) guarantees
that
$\ddot{a}(t)>0$
at least for some time even if a scenario is designed to avoid a
large number of e-foldings.

For a general action, one can define two further dynamical parameters in addition to $\ep$ that (as will be shown in the next section) control the slow roll regime and are directly measurable by observations:
\bea
\obseta&\equiv & \frac{\dot\ep}{H\ep}
\\
\obskappa &\equiv&\frac{\dot{c_s}}{Hc_{s}}.
\eea
These parameters are independent of the choice of scalar field definition (the field gauge choice)  in the action.

\section{slow roll conditions for a general action}\label{sr}
In the case of a general inflation model the term ``slow roll" can be
ambiguous. Here we ensure that slow roll is independent of a scalar field definition, and purely relates to the gauge invariant flow parameters:

\bea \ep, \obseta, \obskappa,   \obseta_N, \obskappa_N... \ll 1,
\label{slowrollconditions}
\eea
where $\obseta_N\equiv
d\obseta/dN_{e}$ etc..
These parameters are dependent upon $\L$ and gauge invariant combinations of $X$ and derivatives of $\L$ with respect to $X$ and $\phi$.

In this paper we do not establish whether particular actions are able
to realize slow roll inflation. However, in the following sections, we
do consider the implications for evolution if slow roll behavior is
satisfied.  From Eq.~(\ref{epsilon}) we can see that $\ep\ll 1$ implies,
\bea
\frac{X\Lx}{-\L} \ll 1. \label{srcond}
\eea
Combined $\ep\ll1$
and $\obseta\ll 1$ suggest that
\bea
\obseta-2\ep&=&{\dot{X\Lx}\over H
X\Lx} \ll 1,
\eea
while $\kappa\ll 1$ implies,
\bea
\frac{1-c_s^{2}}{2}\left|\frac{\dot{(X^2\Lxx)}}{HX^2\Lxx} -\frac{\dot{(X\Lx)}}{HX\Lx} \right|\ll 1.
\eea
Note that, as such, this `slow roll' condition allows for scenarios such as ultra-relativistic DBI inflation in which $c_s^2\ll 1$ if
\bea
\left|\frac{\dot{(X^2\Lxx)}}{HX^2\Lxx} \right|\ll 1.
\eea

\section{Observational constraints on slow roll parameters from the primordial spectrum}\label{pert}

In the absence of anisotropic stress in energy momentum
tensor at linear order,  we can write the metric in the longitudinal
gauge as \cite{Mukhanov:1990me}:
\begin{equation}
ds^2=(1+2\Phi)dt^2-(1-2\Phi)a^2(t)\gamma_{ij}dx^idx^j.
\end{equation}

   Just as in the standard canonical action, we
can define the Bardeen parameter $\zeta$\footnote{Note that the sign
  of $\zeta$ here is the opposite of the sign convention used in
  Section \ref{kinetic}.},
\bea
\zeta&\equiv&{5\rho+3p\over3(\rho+p)}\Phi+{2\rho\over
3(\rho+p)}{\dot{\Phi}\over H},
\label{eq:zetadefforrecons}
\eea
 and Mukhanov variable $\nu$,
 \bea
u &\equiv& z \zeta,
\eea
where for the general action \cite{Garriga:1999vw},
\bea
z&\equiv& {a(\rho+p)^{1/2}\over c_s H},
\\
&=& \frac{\sqrt{2}M_{pl} a\sqrt{\epsilon}}{c_s}.
\eea

In a flat universe, after
quantization, a general action still  has an equation of motion similar to that for
canonical actions \cite{Garriga:1999vw}
\begin{equation}\label{Perteom}
\frac{d^2u_k}{d\tau^2}+\left(c_s^{2}k^2 -\frac{1}{z}\frac{d^2z}{d\tau^2}\right)u_k=0,
\end{equation}
where \bea \frac{1}{z}\frac{d^2z}{d\tau^2} &=& a^{2}H^{2}W, \eea with,
\bea W&=& 2\left[\left(1+\frac{\obseta}{2}-\obskappa\right)
\left(1-\frac{\epsilon}{2}+\frac{\obseta}{4}-\frac{\obskappa}{2}\right)\right]
\nonumber \\ &&+\frac{\obseta_N}{2}-\obskappa_N.  \eea

Now inserting slow roll conditions Eq.~(\ref{slowrollconditions}),
more specifically assuming $\obseta\ll {1\over N}$,
$\ep$ varies very slowly, then using Eq.~(\ref{accel}), $
aH\tau(1-\ep)\approx -1$, and $u_k$ satisfies a Bessel equation,
\bea \frac{d^{2}u_k}{d\tau^2}
+\left(c_s^2k^2-\frac{\nu^2-\frac{1}{4}}{\tau^2}\right)u_k =
0,\label{uofk} \eea where \bea \nu^2
&=&\frac{W}{(1-\ep)^2}+\frac{1}{4}. \eea
In the slow roll limit Eq.~(\ref{slowrollconditions}), solution tends toward
$\nu\rightarrow 3/2$. Following
\citep{Garriga:1999vw}, to leading order the scalar spectral density, ${\cal P}_{\cal
R}$ is given by

\bea {\cal P}_{\cal R}
&=&\frac{k^3}{2\pi^2}\left.\frac{|u_{k}|^2}{z^2}\right|_{c_sk=aH}, \\
&\sim& {1\over 8 \pi^2\mpl^2} {H^2\over c_s \epsilon}|_{c_sk=aH}.  \eea 

The tensor spectra density to
first order is \bea {\cal P}_{h} &=&
\left.\frac{2H^{2}}{\pi^2\mpl^2}\right|_{k=aH}.  \eea Note that in
these computations, we are implicitly assuming Bunch-Davies vacuum
boundary conditions, whose validity generically has model-dependent
limitations \cite{Kinney:2007ag,Chung:2003wn}.  Scalar perturbations
are calculated at sound horizon crossing, $k_s= aH/c_s$, while tensor
perturbations are fixed when $k_t=aH$, so that \bea \left.\frac{d\ln
k}{dN_e}\right|_{k=k_s} &=& -(1-\ep -\obskappa), \\ \left.\frac{d\ln
k}{dN_e}\right|_{k=k_t} &=& -(1-\ep ).  \eea The scalar spectral index
is given by \bea n_s-1 &\equiv&\left. \frac{d\ln {\cal P}_{\cal R}
}{d\ln k}\right|_{k=k_s}, \\
&\approx&-(2\ep+\obseta+\obskappa) +O(\ep^2,\ep\obseta,\obskappa_N,...) , \ \ \ \ \
\ \\ n_t &\equiv& \left.\frac{d\ln {\cal P}_{h} }{d\ln
k}\right|_{k=k_t} \approx- 2\ep+O(\ep^2,...), \eea and tensor to
scalar ratio \bea r\equiv \frac{{\cal P}_{h}}{{\cal P}_{\cal R}}
\approx 16c_s\ep\label{ratio}, \eea gives rise to the consistency
relation \bea r \approx -8c_sn_t. \label{consistency} \eea
Note that the consistency relationship is very similar to that of
multifield models.  The running in the spectral indices are given by
\bea \left.\frac{dn_s}{d\ln k}\right|_{k=k_s} &\approx&
2\ep_N+\obseta_N+\obskappa_N,
\\
\left.\frac{dn_t}{d\ln k}\right|_{k=k_t} &\approx& 2\ep_N.
\eea
As discussed in \cite{Garriga:1999vw}, the consistency relation
could allow us to
determine wether the action is canonical or not in the context of
single field inflation.  Some data fitting results exploring the effects of
including $c_s$ in Eq.\ (\ref{consistency}) can be found in
\cite{Finelli:2006fi}.

 The primordial spectrum, as we will discuss, in theory provides
 information with which we might differentiate between different kinetic inflationary
 models in the slow roll regime defined in Eq.~(\ref{slowrollconditions}).
To obtain this however, requires the
 measurement of both the tensor and scalar primordial spectra,
 including scale dependency of tensor modes, and constraints on
 non-Gaussianity: 1) A measurement of $n_t$ would give a direct
 estimate of $\ep$. 2) Comparing measurements of $r$ and $n_t$ in
 Eq.~(\ref{consistency}) gives a measure of $c_s$ and a first insight into
 whether the action is canonical or not. 3) $n_s$ allows us to
 constrain $2\ep-\obseta-\obskappa$, while $dn_t/d\ln k$, using
 Eq.~(\ref{epN}) would constrain $2\ep-\obseta$. Comparing these two would
 give a direct measure of $\obskappa$. We exhaust the information
 coming from the two-point correlations
since running in the scalar spectral index is dependent on higher
order terms.

Observational tests of non-Gaussianity using the CMB provide
additional measurements of the flow parameters. The CMB is most
sensitive to the 3-point function of the comoving curvature
perturbation, $\zeta$, \bea \zeta=\zeta_G -
\frac{3}{5}f_{NL}\zeta_{G}^{2}, \eea where $\zeta_G$ is a Gaussian
field and where $f_{NL}$ gives a measure of the local intrinsic
non-linearity in the curvature fluctuation as discussed in section
\ref{kinetic}. Non-Gaussianity in general actions has been computed by
\citep{Seery:2005wm,Chen:2006nt} with the definition of $\zeta$ having
the opposite of the sign convention of
Eq. (\ref{eq:zetadefforrecons}).  For generalized single field
inflation, \cite{Chen:2006nt} finds, in the equilateral momentum
triangle limit, \bea
f_{NL}^{\mbox{equil}}&\approx&(-0.26+0.12c_s^2)\left(1-\frac{1}{c_s^2}\right)\nonumber\\&&-0.08\left({c_s^2\over\ep}\right)\frac{X^3\L_{XXX}}{M_{pl}^2H^2},
\ \ \ \ \ \ \eea whereas in the squeezed triangle limit
\cite{Creminelli:2004yq},
\begin{equation}
f_{NL}^{local} \sim (n_s -1).
\end{equation}
Note that the $f_{NL}^{local}$ detection recently reported in
\cite{Yadav:2007yy} uses a sign convention opposite to that used in
\citep{Seery:2005wm,Chen:2006nt}.  The amplitude of the primordial
non-Gaussianity $f_{NL}$ therefore could in principle be used in addition to the scalar and tensor power
spectrum measurements to obtain information about a higher derivative term, $\L_{XXX}$  if one assumes single field inflation.

Current non-Gaussianity limits coming from the equilateral triangle
limit of bispectrum are not competitive with the 2-point constraints,
with WMAP 3-year data giving $-256<f_{NL}^{\mbox{equil}}<332$ at the 95\%
confidence level \citep{Creminelli:2006rz}. Prospectively the PLANCK
satellite will improve this constraint with estimated errors in
$\sigma(f_{NL}^{\mbox{equil}})=66.9$ at 1$\sigma$ level \citep{Smith:2006ud}.
It is however intriguing that \cite{Yadav:2007yy} has very recently
reported a positive dection of local $f_{NL}^{local}$.  Given that
this result came out after our work was completed, we leave the full
discussion of its implications to a future work.  A related discussion
in the context of curvaton models has already appeared \cite{Li:2008qc}.

\section{Reconstructing the action from slow roll parameters} \label{srrecon}

In section \ref{pert} we outlined how power spectrum observations can
give us constraints on the slow roll parameters and $c_s$, which in
their own right can help us differentiate broadly between theories
with $c_s=1$ from $c_s\neq 1$.  In this section we take the slow roll
constraints one step further and consider the question of given the
constraints on these parameters and $c_s$, how much can be known about
the original action of the inflaton field. Although we are only focusing on slow roll parameters and $c_s$, the formalism we discuss in this section can easily be extended to obtain more details about the original action if we include measurement of higher order correlation functions such as $f_{NL}$ which as we explained in the last section contain higher derivative terms.

We consider the `ideal' analytical reconstruction possible if
$\epsilon(k)$ and $c_s(k)$  are measured over some observable range
$k_{min}(N_{e,max})\le k\le k_{max}(N_{e,min})$.

One can reconstruct the evolutionary trajectory for the homogenous
energy density and pressure ($p=\L$)
relative to some reference point within that range, $k_{0}(N_{e0})$.
Combining the definition of $\epsilon$ in Eq.~(\ref{epsilonH}) and the
conservation of energy conservation, Eq.~(\ref{cont}), one finds
\bea
\frac{\rho(N_e)}{\rho(N_{e0})} &=&
\exp\int_{N_{e0}}^{N_e}2\ep(N) dN. \label{rhone}
\eea
where $\rho(N_e)$ and $\L(N_e)$ are related at each time by
\bea
\L(N_e)&=&\rho(N_e) \left(-1+\frac{2\ep(N_e)}{3}\right), \label{Lofepsilon}
\eea

Eq.~(\ref{Lofepsilon}) and using Eq.~(\ref{epsilon}) and Eq.~(\ref{defcs}) give constraints on gauge independent combinations of $X$ and derivatives of $\L$ for the on-shell trajectory,
\bea
X(N_e)\Lx (N_e)&=& \frac{1}{3}\ep(N_e)\rho(N_e) \label{XLx}
\\
X^2(N_e)\Lxx(N_e) &=&\frac{1}{6}\left(\frac{1}{c_s^{2}(N_e)}-1\right)\ep(N_e)\rho(N_e)  \label{X2Lxx} \ \ \ \ \ \
\eea

Notice that here the gauge ambiguity for $X(N_e)$ arises since unlike
the canonical action, not having fixed the kinetic term in the action
and not knowing $\Lx$ as a specific function of $\phi$ or $X$ to
substitute for, leaves the door open to gauge ambiguities due to field
redefinitions, $\phi \rightarrow \varphi \equiv f(\phi)$.  We discuss
the possibility of more general canonical transformations in the
appendix.  Typically, only $\phi \rightarrow \varphi \equiv f(\phi)$
will lead to a local field theory and more general canonical
transformations will not lead to a local theory.  Hence in order to
establish the evolution more specifically we must choose a scalar
field gauge. This can, for example, be done by choosing $\Lx(N_e) =
c_s^{-1}(N_e)$, as is the case for canonical and DBI inflation,
leading to \bea X(N_e)&= &\frac{1}{3}\ep(N_e)c_s(N_e)\rho(N_e)
\label{px} \\ \Delta\phi &\equiv&
\phi(N_e)-\phi(N_{e0})=-\int_{N_{e0}}^{N_{e}}\sqrt{2\epsilon c_s} dN,\
\ \ \ \ \eea where for simplicity in notation for the rest of this
section only, we are using the $M_{pl}=1$ convention (geometricized
units). An alternative useful gauge is taking $X(N_e)=1/2$, for which,
\begin{equation}
\Delta\phi=-\int_{N_{e0}}^{N_{e}} \frac{dN}{H}=-\int_{N_{e0}}^{N_{e}} dN\sqrt{\frac{3}{\rho(N_e)}}~,
\end{equation}

 Notice that here we are aiming to reconstruct a two dimensional
manifold $\L(X, \phi)$ in a three dimensional space $(\L, X, \phi)$
and we have so far shown that after fixing the gauge ambiguity, the
one dimensional trajectory of  $\L({1\over 2},\phi)$ is required to
lie on this manifold and locally minimize it at the same time, however in
the $X$ direction only the first and second derivatives  are
constrained leaving the higher derivatives along $X$ completely
free. Therefore any action consistent with such observations
automatically equates to satisfying the above boundary conditions.
One can easily find all such manifolds of $\L(X,\phi)$ by solving an
arbitrary third order differential equation along characteristic
curves of $(\L~,X~,~\phi=constant)$ obeying the boundary conditions
at $({1\over 2},\phi)$. Any action consistent with the constraints
on $X\Lx$ and $X^2\Lxx$ can then be written in the form
\bea
\label{generalaction} \tilde{\L} &=& q(X,\phi)+\L\left({1\over
2},\phi \right)-q\left({1\over 2},\phi \right)\nonumber
\\
&+&
\left[\Lx \left({1\over 2},\phi \right)-q_{X} \left({1\over 2},\phi \right)\right] \left(X-{1\over
2} \right)\nonumber
\\
&+&{1\over 2}\left[\Lxx \left({1\over2},\phi \right)-q_{XX}
\left({1\over 2},\phi \right)\right] \left(X-{1\over2} \right)^2 \ \
\ \
\eea
where $q$ is an arbitrary function of $\phi$ and $X$. It is
also straight forward to  show that the trajectory of
$\tilde{\L}({1\over 2},\phi)$ is minimizing the action, since
equation of motion Eq.~(\ref{eom}) for $\tilde{\L}$ at $X={1/2}$
simplifies to,
\bea
\Lp \left({1\over 2},\phi \right)-
\Lxp\left({1\over 2},\phi \right) = 3H(N_e)\Lx\left({1\over 2},\phi
\right),
\eea which, using  Eq.~(\ref{XLx}), turns up to be simply an
alternative way of writing $\rho_N = 2\ep \rho$ which has already
been set to remain valid. This can be seen more clearly through the
following example. Lets consider the case where  $\epsilon \sim
{1\over 2 N_e}\ll 1$ where we are taking $N_e$ to be decreasing
during inflation. This is what one would expect for a quadratic
potential in the case of a canonical action $c_s=1$. We will consider the implications for the action if $c_s$ deviates slightly from one, $c_s=1-\delta$. We first obtain $H$ using Eq.~(\ref{rhone}):
\bea
H=H_1\exp \left(\int^{N_e}_1 {dN\over 2N}\right)=H_1N_e^{1/2}
\eea
 where
$H_1=H|_{N_e=1}$ . Now fixing the gauge to $X=1/2$ we get
\bea
{d\phi\over dN}={-1\over H_1N_e^{1/2}}  \Rightarrow \phi =- 2
{N_e^{1/2}\over H_1}
\eea
the above equation combined with
Eq.~(\ref{XLx}) and Eq.~(\ref{X2Lxx}) yield
\bea
\L\left({1\over 2},\phi \right)&=&H_1^2(1-H_1^2\phi^2) \\
\Lx \left({1\over 2},\phi \right)&=&H_1^2 \\
\Lxx \left({1\over2},\phi \right)&\sim&2H_1^2\delta
\eea
Now substituting these result in Eq.~(\ref{generalaction}), and for
instance taking $q=0$, in the limit of $\epsilon\ll1$ or equivalently
$|H_1\phi|\gg1$ the action will have a following form:
\bea
\tilde{\L}_1(X,\phi)\sim H_1^2\left[-{3\over
4}(H_1\phi)^2+X+\delta ~X^2\right],
 \eea which after a field
redefinition is slightly deviated from the a canonical action with
quadratic potential. However if we take $q=\lambda X^3$ then the
action will be:
\bea
\tilde{\L}_2(X,\phi)&=&\tilde{\L}_1(X,\phi)\nonumber\\&+&\lambda\left[X^3-{1\over
8}-{3\over 4}(X-{1\over 2})-{3\over
2}(X-{1\over2})^2\right]\nonumber
\\
\eea which also satisfies the
equation of motion at $X={1\over 2}$ and fits $\epsilon$ and $c_s$
regardless of the magnitude of $\lambda$.
\section{Inflationary flow equations}\label{flow}

In the previous section it was shown that,
even after fixing a gauge, there are an infinite number of
different actions that can match the same observation,  however it is
possible to write down one dynamical evolution for the on-shell
trajectory for all of them. That is to say, just like the canonical
case, we can obtain $H(\phi)$ or the function $\L(\phi)$ on the
solution trajectory but, unlike before where it would be equated to a
unique potential $V(\phi)$, it will not correspond to a unique
$\L(X,\phi)$.

mAll the gauge invariant parameters that we have introduced so far
belong to two categories: first, combinations of $H$ and its
derivatives with respect to e-folding number $N_e$: \bea
H&, &~ \epsilon={d\ln H\over dN_e},  \nonumber \\
\eta&=&-{d\ln \epsilon\over dN_e}={1\over \epsilon} {d^2 \over
dN_e}\ln H, ~ \eta_N , ~\eta_{NN}, ... \eea and second, combinations
of $c_s$ and its derivatives with respect to $N$: \bea c_s, ~
\obskappa={1\over c_s} {d c_s\over dN_e}, ~~ \obskappa_N, ...~~~~~
\eea
By truncating these parameters at some derivative order to zero and
then setting initial values for rest of them at $N_e=N_{e0}$ one
could approximate  $H$ or $c_s$ with Taylor expansions in terms of
$N_e$ up to a convergence radius $N_{max}$.

An alternative approach using {\it inflationary flow equations} to describe an action beyond the slow-roll assumption has been used extensively for canonical inflation \citep{Liddle:1994dx,Kinney:2002qn,Liddle:2003py} and DBI inflation \citep{Peiris:2007gz}. Here we discuss how this formalism can be extended to a general action and a general gauge.

The inflationary flow equations are used  to derive a Taylor expansion of $H$, $\L$, and $c_s$ and other gauge invariant quantities in terms of  a specific choice of scalar field, $\phi$, for example
\bea
H(\phi)& =&H_0+ M_{pl}H^\prime_0\left(\phimp\right) +.... \nonumber\\
&+&{1\over(
l+1)!}M_{pl}^{l+1}H^{[l+1]}_0\left(\phimp\right)^{l+1}+...~,
\eea
and hence the coefficients have the nontrivial terms in the form of
,
\bea
Q_{l}(H)|_{\phi_0}
&=&
\left[\left({dN_e\over d\phi}{d\over dN_e}\right)^{l}H\right]
_{\phi_0}\ \ \ \
\eea
and similarly terms of the form $Q_l(c_s)$ for the Taylor expansion of $c_s$.
Since $X$ and $\Lx$ are not invariant under the
field redefinition and
\bea \label{dphidN} {dN_e\over
d\phi}=\pm {H\over \sqrt{2X}}=\pm \left({\Lx\over 2\ep}\right)^{1/2},
\eea
the $Q_l(H)$ and $Q_l(c_s)$ are in general gauge dependent. Fixing a gauge, as is done in DBI and canonical inflation with  $\Lx=c_s^{-1}$, sets this dependency.

For a general gauge, we  can write the gauge invariant slow roll parameters as
\bea
\epsilon &=& \frac{2\mpl^2}{\Lx} \left(\frac{\Hp}{H}\right)^2,
\\
\obskappa &=& \frac{2\mpl^2 }{\Lx}\left(\frac{\Hp}{H}\frac{(c_s^{-1})'}{c_s^{-1}}\right),
\eea
and introduce gauge dependent parameters
\bea
\gaugeeta &\equiv&\frac{2\mpl^2}{\Lx} \left(\frac{\Hpp}{H}\right) ,
\\
\gaugekappa &\equiv&\frac{2\mpl^2 }{\Lx}\left(\frac{\Hp}{H}\frac{\LxP}{\Lx}\right).
\eea
where in canonical and DBI inflation the gauge choice leads to $\gaugekappa = \obskappa$.

 $\gaugeeta$ and $\gaugekappa$ are not invariant under a redefinition of the scalar field $\phi\rightarrow \varphi(\phi)$,
\bea
\gaugeeta&=&-{\dot{X}\over 2H
X}-{\dot{\Lx}\over H \Lx},
\\
\gaugekappa&=&-{\dot{\Lx}\over H\Lx},
\eea
however, the combination $ 2\gaugeeta-\gaugekappa$ is invariant under the transformation,
 \bea
 2\gaugeeta-\gaugekappa = 2\ep-\obseta =  -\frac{\dot{(X\Lx)}}{HX\Lx}.
  \eea

The inflationary flow equation hierarchy is obtained by defining three sets of variables,
\bea
{^{l}\lambda(\phi)} &\equiv&\left(\frac{2\mpl^2}{\Lx}\right)^{l}\left(\frac{\Hp}{H}\right)^{l-1}\frac{H^{[l+1]} }{H}
\\
{^{l}\flowb(\phi)} &\equiv&\left(\frac{2\mpl^2}{\Lx}\right)^{l}\left(\frac{\Hp}{H}\right)^{l-1}
\frac{(c_s^{-1})^{[l+1]}}{c_s^{-1}}
\\
{^{l}\flowc(\phi)} &\equiv&\left(\frac{2\mpl^2}{\Lx}\right)^{l}\left(\frac{\Hp}{H}\right)^{l-1}
\frac{\Lx^{[l+1]}}{\Lx}
\eea
for $l\ge 1$, where $H^{[l+1]}\equiv d^{l+1}H/d\phi^{l+1}$ and  $\gaugeeta = {^1\lambda}$. For DBI inflation and canonical inflation
${^l\flowb}={^l\flowc}$. As explained above, in general ${^l\lambda}, {^l\flowb}$ and ${^l \flowc}$ are not invariant under scalar field redefinitions.

Noting that
\bea
\frac{d\phi}{dN_e} = \frac{2\mpl^2}{\Lx}\frac{\Hp}{H}\label{phiNe}
\eea
the evolutionary paths of these parameters simplify to coupled first order differential equations with respect to $N_{e}$. Then,
\bea
\ep_N &=& -\ep(2\ep-2\gaugeeta+\gaugekappa) = -\ep\eta, \label{epN}
\\
\gaugeeta_N &=& -\gaugeeta(\ep+\gaugekappa) +{^2\lambda},
\\
\obskappa_N &=& -\obskappa(\ep-\gaugeeta+\gaugekappa+\obskappa)+\ep{^1\flowb},
\\
\gaugekappa_N &=& -\gaugekappa(\ep-\gaugeeta+2\gaugekappa)+\ep{^1\flowc},
\eea
and for $l\ge 1$,
\bea
{^{l}\lambda_N} &=& -^l\lambda\left[l\ep-(l-1)\gaugeeta+l\gaugekappa\right] + {^{l+1}\lambda}, \ \ \ \ \ \ \label{flow1}
\\
{^{l}\flowb_N} &=& -{^{l}\flowb}\left[(l-1)\ep-(l-1)\gaugeeta+l\gaugekappa+\obskappa\right]+{^{l+1}\flowb}, \ \ \ \ \ \ \ \ \label{flow3}
\\
{^{l}\flowc_N} &=& -{^{l}\flowc}\left[(l-1)\ep-(l-1)\gaugeeta+(l+1)\gaugekappa\right]+{^{l+1}\flowc}. \ \ \ \ \ \ \ \ \ \label{flow2}
\eea

Following the nomenclature of \cite{Liddle:2003py},  the Taylor
expansion of the Hubble factor, $c_s^{-1}$ and $\Lx$  in powers of
$\phi$ can be written, 
\bea H(\phi) &=& H_0\left[ 1+
A_1\left(\phimp\right) +.... \right.\nonumber
\\&& \left. \ \ \ \ +A_{M_A+1}\left(\phimp\right)^{M_A+1}+...\right], \ \ \  \label{Hflow}
\\
c_s^{-1}(\phi) &=&  c_{s0}^{-1}\left[ 1+ B_1\left(\phimp\right) + ...\right. \nonumber
\\ && \left. \ \ \ \ +B_{M_B+1}\left(\phimp\right)^{M_B+1}+...\right], \ \ \ \label{csflow}
\\
\Lx(\phi) &=& \Lxzero\left[ 1+ C_1\left(\phimp\right) + ...\right. \nonumber
\\ && \left. \ \ \ \ +C_{M_C+1}\left(\phimp\right)^{M_C+1}+...\right], \ \ \ \label{Lxflow}
\eea
where
\bea
 A_{l} &\equiv &\frac{1}{l!}\left.\frac{{\mpl}^{l}}{H_0} H^{[l+1]}\right|_{\phi=\phi_0}
   \\
  B_{l} &\equiv& \frac{1}{l!}\left.\frac{{\mpl}^{l}}{c_{s0}^{-1}} (c_s^{-1})^{[l+1]}\right|_{\phi=\phi_0}
   \\
  C_{l} &\equiv&\frac{1}{l!} \left.\frac{{\mpl}^{l}}{\Lxzero} \Lx^{[l+1]}\right|_{\phi=\phi_0}
\eea
and $H_{0}$, $\Lxzero $ and $c_{s0}$ are the values of $H$, $\Lx$ and $c_s$ at the reference point $\phi_0\equiv \phi(N_{e0})$ with  $\Delta\phi \equiv\phi(N_e)-\phi(N_{e0})$. Note that for scenarios such as relativistic
DBI where $c_s^{-1}$ diverges at desiter limit and Taylor expansion description is invalid out of the convergence radius of $\phi_0$, instead one could use the Taylor expansion of $c_s$ and $\Lx^{-1}$.

Using Eq.~(\ref{Hprime}) and Eq.~(\ref{flow1}) - Eq.~(\ref{flow3}),
\bea
A_{1} &=&\sqrt{\frac{\ep_0 \Lxzero}{2}},
\\
A_{l+1} &=&\frac{ \Lxzero ^{l}}{2^{l}(l+1)!A_{1}^{l-1}}{^l\lambda_0},
\\
B_{1} &=& \frac{\obskappa_0 \Lxzero}{2A_1},
\\
B_{l+1} &=&\frac{\Lxzero ^{l+1}}{2^{l}(l+1)!A_{1}^{l-1}}{^l\flowb_0}.
\\
C_{1} &=& \frac{\gaugekappa_0 \Lxzero}{2A_1},
\\
C_{l+1} &=&\frac{ \Lxzero ^{l+1}}{2^{l}(l+1)!A_{1}^{l-1}}{^l\flowc_0},
\eea

The flow equations derived in this section
apply to inflationary models independent of whether inflation is slow roll or not.
Often in applying the flow equations, however, it is commonly assumed that within the chosen gauge, the series are convergent,  and the hierarchies in ${^{l}\lambda}$, ${^{l}\alpha}$ and ${^{l}\beta}$ can be truncated with non-zero values for a finite range of $l$, $l\leq M_{A}$, $l\leq M_{B}$ and $l\leq M_C$ respectively in the chosen gauge.

\section{Conclusions}\label{conc}

Complementary CMB and large scale structure measurements over scales
spanning four orders of magnitude have driven impressive improvements
in the measurement of the primordial scalar power spectrum. In
addition improvements in non-Gaussianity constraints are expected from
the PLANCK satellite and there is the exciting prospect of tensor mode
measurements in the near future, with a number of CMB surveys being
developed to target B-mode polarization.  With the hope of connecting
this to high energy theory, there has been significant interest in
establishing what current and planned observations might elucidate
about the primordial spectrum of fluctuations from inflation, in terms
of potential reconstruction for canonical inflation and action
reconstruction in specific theories such as DBI inflation.

In this paper we extend these considerations to address what we can
maximally learn about the inflationary action without making any
assumptions, a priori, about its form.  We establish how observational
constraints on the inflationary slow roll parameters could be
successfully applied to reconstruct the general action over observable
scales in the context of single field inflationary models.  Under the
assumption of slow roll inflation, we have demonstrated that in an
idealized case in which $\{H, c_s, \ep,\obseta,\obskappa\}$ are measured over
a finite range of scales, we
analytically obtain the trajectory of the general action $\L, X\Lx,
X^2\Lxx$,  independent of the scalar field definition, with
respect to some reference point. With the specification of a gauge,
the measurement of the first level of flow parameters enables trajectories
of $X, \Lx$ and $\Lxx$ and information about $\Lp$ and $\Lxp$ to be
established.

Using the Hamilton-Jacobi formalism, we extend the inflationary flow
parameter approach to describe the evolutionary trajectories of
general actions. This involves introducing three hierarchies of flow
parameters to describe the evolution of a general action without
using the specific gauge, $\Lx=c_s^{-1}$, used in
canonical and DBI inflation. These equations hold for all single field
inflationary scenarios, whether or not slow roll conditions are met.

Observations promise to allow us to reconstruct a wealth of
information about the general action including powerful insights into
the form of the Lagrangian kinetic, potential and hybrid terms and the
relative importance of kinetic and potential components over the
course of the trajectory.
As it is difficult to obtain large observable non-Gaussianities
without non-minimal kinetic terms (and/or resorting to
curvaton scenarios), future detection of non-Gaussianities would make
formalisms such as the one we present here indispensable to understand
what kind of high energy theories are compatible with cosmological
data.
 This is good news for PLANCK and other future CMB
experiments which are certain to obtain increasingly precise data regarding
non-Gaussianities, and it is also good news for high energy theorists
looking for distinctive clues to the identity of the inflaton.
 In work in preparation, we are investigating the observational
constraints on the general inflationary action using this formalism.

\section*{Acknowledgements}
The authors would like to thank Niayesh Afshordi, Nishant Agarwal, Lisa Everett and Gary Shiu for helpful comments and discussions.
The work of RB is supported by the National Science Foundation under
grants AST-0607018 and PHY-0555216. The work of DJHC and GG is
supported by the DOE Outstanding Junior Investigator Program through grant
DE-FG02-95ER40896.  The work of GG was also supported by Perimeter Institute for Theoretical Physics.  Research at Perimeter Institute is supported by the Government of Canada through Industry Canada and by the Province of Ontario through the Ministry of Research \& Innovation.
\appendix
\section{Canonical Transformations versus Field Redefinitions}
Although a local field redefinition of the form
$\phi=\phi(\tilde{\phi}$) yields a classical Lagrangian density which
describes the same physics, such transformations only form a subset of
the canonical transformations $\phi=\phi(\tilde{\phi},\tilde{\pi})$
and $\pi=\pi(\tilde{\phi},\tilde{\pi})$ which by construction
preserves the physics. In this Appendix, it is shown how the
Lagrangian transforms under a more general set of canonical
transformations.

Restricting to 0+1 dimensions, we
find the interesting result that a minimal kinetic term can be
transformed into a system with non-minimal kinetic term (non-minimal here is to be distinguished from
non-canonical since the latter is a straightforward transformation). As a
byproduct, we find an exact solution to the non-linear differential
equation Eq.~(\ref{eq:origvareq}) through the use of a canonical
transformation. Unfortunately, the generalization of such minimal to
non-minimal transitioning systems to 3+1 dimensions results in a
non-local Lagrangian.

To begin with, let us show that a local field redefinition of the
form $\phi=g(\tilde{\phi})$ leads to a physically equivalent equation
of motion. Consider a Lagrangian density of the form $\mathcal{L}(X,\phi)$
where $X\equiv(\partial\phi)^{2}$. Taking the variation of the action\begin{equation}
S=\int d^{4}x\mathcal{L}(X,\phi)\end{equation}
 yields the EOM \begin{equation}
2\partial_{\mu}\{\partial^{\mu}\phi\frac{\partial}{\partial X}\mathcal{L}(X,\phi)\}-\frac{\partial\mathcal{L}}{\partial\phi}=0.\end{equation}
 Define the local field redefinition\begin{equation}
\phi=g(\tilde{\phi}).\end{equation}
 We then have
\bea
&& 2g'(\tilde{\phi})\partial_{\mu}\left\{\frac{1}{g'(\tilde{\phi})}\right\}\partial^{\mu}\tilde{\phi}\frac{\partial\mathcal{L}((g'(\tilde{\phi}))^{2}\tilde{X},g(\tilde{\phi}))}{\partial\tilde{X}}-\frac{\partial\mathcal{L}}{\partial\tilde{\phi}} \nonumber
\\
&&+2\partial_{\mu}\left\{\partial^{\mu}\tilde{\phi}\frac{\partial\mathcal{L}((g'(\tilde{\phi}))^{2}\tilde{X},g(\tilde{\phi}))}{\partial\tilde{X}}]\right\}+2g''\frac{\tilde{X}}{g'}\frac{\partial\mathcal{L}}{\partial\tilde{X}}=0. \nonumber
\\ &&
\eea
Because the first and the last terms cancel, we end up with an equation
of motion for a new Lagrangian of the form\begin{equation}
\tilde{\mathcal{L}}(\tilde{X},\tilde{\phi})=\mathcal{L}((g'(\tilde{\phi}))^{2}\tilde{X},g(\tilde{\phi})).
\end{equation}

The stress energy tensor for the new Lagrangian density can also be
checked to be physically identical to the original:\begin{equation}
\tilde{X}=g^{\mu\nu}\partial_{\mu}\tilde{\phi}\partial_{\nu}\tilde{\phi}\end{equation}
\begin{equation}
S=\int d^{4}x\sqrt{g}\tilde{\mathcal{L}}\end{equation}
\begin{equation}
T_{\mu\nu} =  \frac{2}{\sqrt{g}}\frac{\delta S}{\delta g^{\mu\nu}}.\end{equation}
Hence, it is clear that a local field redefinition leads to the same
physics.  Now, let us consider the more general possibility of a
canonical transformation.

Restrict to the 0+1 dimension inflaton theory, which would
correspond to a classical mechanics problem in one spatial dimension.
A transformation from the phase space variable $\{\phi,p\}$ to $\{\tilde{\phi},\tilde{p}\}$

\begin{equation}
\phi=\phi(\tilde{\phi},\tilde{p};t)\end{equation}
\begin{equation}
p=p(\tilde{\phi},\tilde{p};t)\end{equation}
corresponds to a canonical transformation if Hamilton's equations
are preserved, which in turn implies\begin{equation}
p\dot{\phi}-H(\phi,p,t)=\tilde{p}\dot{\tilde{\phi}}-\tilde{H}(\tilde{\phi},\tilde{p},t)+\frac{d}{dt}F(\phi,\tilde{\phi},t)\end{equation}
for some function $\tilde{H}$ and $F$. The function $F(\phi,\tilde{\phi},t)$
is called the generating function for the canonical transformation.
The canonical transformation generated by $F$ is then \begin{equation}
p=\frac{\partial}{\partial\phi}F(\phi,\tilde{\phi},t)\end{equation}
\begin{equation}
-\tilde{p}=\frac{\partial}{\partial\tilde{\phi}}F(\phi,\tilde{\phi},t)\end{equation}
with the new Hamiltonian given by\begin{equation}
\tilde{H}(\tilde{\phi},\tilde{p},t)=H(\phi,p,t)+\frac{\partial}{\partial t}F(\phi,\tilde{\phi},t).\label{newh}\end{equation}

Since the Lagrangian is a Legendre transformation of the Hamiltonian,
we have\begin{equation}
\tilde{L}(\tilde{\phi},\dot{\tilde{\phi}};t)=\tilde{p}\dot{\tilde{\phi}}-\tilde{H}(\tilde{\phi},\tilde{p},t)\end{equation}
where \begin{equation}
\dot{\tilde{\phi}}=\frac{\partial\tilde{H}}{\partial\tilde{p}}.\end{equation}
This is the new Lagrangian generated by a canonical transformation,
which contains the same physics.
For example, as long as the canonical transformation is accomplished in
a time independent manner, the energy density remains the same since
$H=\tilde{H}$ according to Eq.~(\ref{newh}).

One may try to express $\tilde{L}$ more directly in terms of $F$
by formally solving some of the algebraic relations above, but as
we will see the final result is not that illuminating except for seeing
how the generating function explicitly mixes $\phi$ and $p$ in the
field redefinition. Start with the Hamiltonian after the canonical
transformation written as

\bea
&& \tilde{H}(\tilde{\phi},\tilde{p},t)= \nonumber
\\ && \ H\left(\phi=\phi_{*}(\tilde{\phi},\tilde{p},t),\frac{\partial}{\partial\phi}F(\phi,\tilde{\phi},t)|_{\phi=\phi_{*}(\tilde{\phi},\tilde{p},t)},t\right) \nonumber
\\ && \  +\frac{\partial}{\partial t}F(\phi=\phi_{*}(\tilde{\phi},\tilde{p},t),\tilde{\phi},t)
\eea
where $\phi_{*}$ solves the equation\begin{equation}
-\tilde{p}=\frac{\partial}{\partial\tilde{\phi}}F(\phi,\tilde{\phi},t)|_{\phi=\phi_{*}}.\end{equation}
Note that this amounts to a field redefinition involving both $\phi$
and $p$. Hence, the Lagrangian becomes
\begin{equation}
\tilde{L}(\tilde{\phi},\dot{\tilde{\phi}};t)=\tilde{p}\dot{\tilde{\phi}}-\tilde{H}(\tilde{\phi},\tilde{p},t)
\end{equation}
where $\tilde{p}$ is eliminated by solving the equation
\bea
\dot{\tilde{\phi}} & = & \frac{\partial\tilde{H}}{\partial\tilde{p}}
\\
 & = & \frac{\partial}{\partial\tilde{p}}\left[H(\phi=\phi_{*}(\tilde{\phi},\tilde{p},t),\frac{\partial}{\partial\phi}F(\phi,\tilde{\phi},t)|_{\phi=\phi_{*}(\tilde{\phi},\tilde{p},t)},t)\right. \nonumber
 \\
 && \ \left.+\frac{\partial}{\partial t}F(\phi=\phi_{*}(\tilde{\phi},\tilde{p},t),\tilde{\phi},t)\right]
 \eea
Unfortunately, there does not seem to be an elucidating general simplification
for $\tilde{L}$. Hence, we turn to some explicit examples.

Consider the original Lagrangian to be\begin{equation}
L=\frac{1}{2}\dot{\phi}^{2}-\frac{1}{2}m^{2}\phi^{2}\end{equation}
and the generating function\begin{equation}
F(\phi,\tilde{\phi},t)=\phi\tilde{\phi}^{2}.\label{eq:firstgenerator}\end{equation}
The Hamiltonian can be obtained as follows:\begin{equation}
p=\dot{\phi}\end{equation}
\bea
H(\phi,p) & = & p\dot{\phi}-L\\
 & = & \frac{1}{2}(p^{2}+m^{2}\phi^{2})\eea
\begin{equation}
p=\tilde{\phi}^{2}\end{equation}
\begin{equation}
-\tilde{p}=2\phi\tilde{\phi}\end{equation}
\bea
\tilde{H} & = & \frac{1}{2}(p^{2}+m^{2}\phi^{2})\\
 & = & \frac{1}{2}(\frac{m^{2}}{4\tilde{\phi}^{2}}\tilde{p}^{2}+\tilde{\phi}^{4})\eea
Hence, we have\begin{equation}
\dot{\tilde{\phi}}=\frac{m^{2}}{4\tilde{\phi}^{2}}\tilde{p}\end{equation}
\bea
\tilde{L} & = & \tilde{p}\dot{\tilde{\phi}}-\frac{1}{2}(\frac{m^{2}}{4\tilde{\phi}^{2}}\tilde{p}^{2}+\tilde{\phi}^{4})\\
 & = & \frac{1}{m^{2}}[2\tilde{\phi}^{2}\dot{\tilde{\phi}}^{2}-\frac{m^{2}}{2}\tilde{\phi}^{4}]\eea
In this case, the field redefinition $\phi=\tilde{\phi}^{2}$ would
have generated the equivalent Lagrangian.

Next, we consider an example in which a non-minimal kinetic term can
be transformed into a minimal kinetic term.
Consider the original Lagrangian to be\begin{equation}
L=\frac{1}{2}\dot{\phi}^{2}-V(\phi)\label{eq:canonicalkinlag}\end{equation}
and the generating function\begin{equation}
F(\phi,\tilde{\phi},t)=\phi\tilde{\phi}^{2}+\tilde{\phi}^{3}.\label{eq:secondgenerator}\end{equation}
The Hamiltonian can be obtained as follows:\begin{equation}
p=\dot{\phi}\label{eq:pandphidot}\end{equation}
\bea
H(\phi,p)
 & = & \frac{1}{2}p^{2}+V(\phi)\eea
\begin{equation}
p=\tilde{\phi}^{2}\label{eq:pandphitilde}\end{equation}
\begin{equation}
-\tilde{p}=2\phi\tilde{\phi}+3\tilde{\phi}^{2}\label{eq:ptildeeq}\end{equation}
\bea
\tilde{H}
 & = & \frac{1}{2}\tilde{\phi}^{4}+V(\frac{-\tilde{p}-3\tilde{\phi}^{2}}{2\tilde{\phi}})\eea
Hence, we have\begin{equation}
\dot{\tilde{\phi}}=\frac{-1}{2\tilde{\phi}}V'(\frac{-\tilde{p}-3\tilde{\phi}^{2}}{2\tilde{\phi}})\end{equation}
which allows us to express $\tilde{p}$ in terms of $\tilde{\phi}$
and $\dot{\tilde{\phi}}$:\begin{equation}
\tilde{p}=-2\tilde{\phi}V'^{-1}(-2\tilde{\phi}\dot{\tilde{\phi})}-3\tilde{\phi}^{2}\end{equation}
 Hence, our Lagrangian becomes\bea
\tilde{L} & = & \tilde{p}\dot{\tilde{\phi}}-\left[\frac{1}{2}\tilde{\phi}^{4}+V\left(\frac{-\tilde{p}-3\tilde{\phi}^{2}}{2\tilde{\phi}}\right)\right]
\\
 & = & -2\tilde{\phi}\dot{\tilde{\phi}}V'^{-1}(-2\tilde{\phi}\dot{\tilde{\phi})}-\frac{d}{dt}\tilde{\phi}^{3}\nonumber
 \\
 && -\left[\frac{1}{2}\tilde{\phi}^{4}+V(V'^{-1}(-2\tilde{\phi}\dot{\tilde{\phi})})\right]\eea

Suppose we consider
$V=\frac{1}{4}\lambda\phi^{4}$. We would find\begin{equation}
V'(\phi)=\lambda\phi^{3}\end{equation}
giving \begin{equation}
V'^{-1}(-2\tilde{\phi}\dot{\tilde{\phi})}=\frac{1}{\lambda^{1/3}}(-2\tilde{\phi}\dot{\tilde{\phi})}^{1/3}\end{equation}
\bea
V(V'^{-1}(-2\tilde{\phi}\dot{\tilde{\phi})})
 & = & \frac{1}{4}\frac{1}{\lambda^{1/3}}(-2\tilde{\phi}\dot{\tilde{\phi})}^{4/3}\eea
\bea
\tilde{L}
 & = & \frac{3}{4}\frac{1}{\lambda^{1/3}}(-2\tilde{\phi}\dot{\tilde{\phi}})^{4/3}-\frac{d}{dt}\tilde{\phi}^{3}-\frac{1}{2}\tilde{\phi}^{4}\label{eq:noncanlagrangian}\eea
Hence, the interesting point of this example is that a canonical transformation
has turned an analytic kinetic term into a non-analytic one.

Let's check that the equation of motion generated from this Lagrangian
can give the same solution as the original Lagrangian. The equation
of motion with this Lagrangian is\begin{equation}
\tilde{\phi}\ddot{\tilde{\phi}}-3(2^{-1/3})\lambda^{1/3}\tilde{\phi}^{8/3}\dot{\tilde{\phi}}^{2/3}-\dot{\tilde{\phi}}^{2}=0.\label{eq:neweom}\end{equation}
To compare to the solutions of the originl equation, \begin{equation}
\ddot{\phi}+\lambda\phi^{3}=0\label{eq:origvareq}\end{equation}
we need to consider an observable and a boundary condition. Since
we are looking at Minkowski physics, we can simply choose the energy
density to be the observable. As far as mapping the boundary conditions
are concerned, note that Eqs.~(\ref{eq:pandphitilde}) and (\ref{eq:pandphidot})
imply\begin{equation}
\tilde{\phi}^{2}(0)=\dot{\phi}(0)\label{eq:bcmap1}\end{equation}

\begin{equation}
2\tilde{\phi}\dot{\tilde{\phi}}=\ddot{\phi}(0)=-\lambda\phi^{3}(0).\label{eq:bcmap2}\end{equation}

Now, the solution to the original variable equation Eq.~(\ref{eq:origvareq})
with the boundary condition \begin{equation}
\phi(t=0)=0\end{equation}
\begin{equation}
\dot{\phi}(t=0)=A\end{equation}
has a solution \begin{equation}
\phi(t)=At[1-\frac{\lambda}{20}A^{2}t^{4}+\mathcal{O}(\lambda^{2}A^{4}t^{8})].\label{eq:leadingordersol}\end{equation}
To compare, using Eqs.~(\ref{eq:bcmap1}) and (\ref{eq:bcmap2}),
we should solve Eq.~(\ref{eq:neweom}) with the boundary conditions\begin{equation}
\tilde{\phi}(0)=\sqrt{A}\end{equation}
\begin{equation}
\dot{\tilde{\phi}}(0)=0.\end{equation}
We see that in fact, in the $\tilde{}$ variables, \begin{equation}
\tilde{\phi}(t)=\sqrt{A}\label{eq:newsol}\end{equation}
 is an exact solution satisfying the desired boundary conditions.

The stress energy tensors to compare are then\begin{equation}
T_{00}=\frac{1}{2}\dot{\phi}^{2}+\frac{\lambda}{4}\phi^{4}\label{eq:originalstress}\end{equation}
and \begin{equation}
\tilde{T}_{00}=\frac{1}{2}\tilde{\phi}^{4}+\frac{1}{4}\frac{1}{\lambda^{1/3}}(-2\tilde{\phi}\dot{\tilde{\phi})}^{4/3}\label{eq:newstress}\end{equation}
Inserting Eq.~(\ref{eq:leadingordersol}) into
Eq.~(\ref{eq:originalstress}), we obtain\begin{equation}
T_{00}\approx\frac{A^{2}}{2}+\mathcal{O}(t^{8})\end{equation} where if
we had not solved the equation of motion, we would have had a $t^{4}$
term on the right hand side. On the other hand, inserting
Eq.~(\ref{eq:newsol}) into Eq.~(\ref{eq:newstress}), we
obtain\begin{equation} \tilde{T}_{00}=\frac{A^{2}}{2}\end{equation}
exactly. Hence, we have given a non-trivial check that the solution
arising from the non-minimal kinetic Lagrangian of
Eq.~(\ref{eq:noncanlagrangian}) gives the exactly the same observable
as the solution arising from the minimal kinetic Lagrangian of
Eq.~(\ref{eq:canonicalkinlag}) with
$V(\phi)=\frac{\lambda}{4}\phi^{4}$.

Thus far, we had been working in 0+1 dimensions (i.e. the spatial
variation of the field had been frozen). Let us consider how this
generalizes to field theory. Unfortunately, we will show that the
interesting example of minimal kinetic term leading to a nonminimal
kinetic term requires a non-local transformation.  First, we would
like to show that
\begin{equation}
\phi=\phi(\tilde{\phi},\tilde{\pi})\end{equation}
\begin{equation}
\pi=\pi(\tilde{\phi},\tilde{\pi})\end{equation}
can be generated by the generating function $F(\phi,\tilde{\phi})$
with the new Hamiltonian given by\begin{equation}
\tilde{\mathcal{H}}(\tilde{\phi},\tilde{\pi})=\mathcal{H}(\phi,\pi)\label{eq:hamiltonianpi}\end{equation}
\begin{equation}
\pi=\frac{\partial}{\partial\phi}F(\phi,\tilde{\phi})\label{eq:pi1}\end{equation}
\begin{equation}
-\tilde{\pi}=\frac{\partial}{\partial\tilde{\phi}}F(\phi,\tilde{\phi}).\label{eq:pi2}\end{equation}
To begin, take the total time derivative of $F$:\begin{equation}
\frac{d}{dt}F(\phi,\tilde{\phi})=\frac{\partial F}{\partial\phi}\dot{\phi}+\frac{\partial F}{\partial\tilde{\phi}}\dot{\tilde{\phi}}.\end{equation}
Using this with Eqs.~(\ref{eq:pi1}) and (\ref{eq:pi2}), we have\begin{equation}
\tilde{\pi}\dot{\tilde{\phi}}+\frac{d}{dt}F(\phi,\tilde{\phi})=\pi\dot{\phi}.\end{equation}
Next, using Eq.~(\ref{eq:hamiltonianpi}), we find \begin{equation}
\tilde{\pi}\dot{\tilde{\phi}}-\tilde{\mathcal{H}}(\tilde{\phi},\tilde{\pi})+\frac{d}{dt}F(\phi,\tilde{\phi})=\pi\dot{\phi}-\mathcal{H}(\phi,\pi),\end{equation}
which says that the two Lagrangian densitites are identical up to
a total time derivative. Note that the total derivative can be non-trivial
when it involves $\phi$ and $\tilde{\phi}$ and not just $\phi$
or $\tilde{\phi}$.

To obtain the new Lagrangian, we use\begin{equation}
\dot{\tilde{\phi}}(x)=\frac{\delta}{\delta\tilde{\pi}(x)}\int
d^{3}x\tilde{\mathcal{H}}\label{eq:pitophi}\end{equation} to solve for
$\tilde{\pi}(x)$. Unfortunately, as we will now show, this method
generally fails to produce a local Lagrangian since solving
Eq.~(\ref{eq:pitophi}) for $\tilde{\pi}$ generically invovles solving
an elliptic PDE.  To see this, start with
\begin{equation}
\mathcal{L}=\frac{1}{2}{(\partial\phi)}^{2}-V(\phi).\end{equation}
The Hamiltonian can be obtained as follows:\begin{equation}
\pi=\dot{\phi}\end{equation}
\bea
\mathcal{H}(\phi,p)
 & = & \frac{1}{2}[\pi^{2}+(\nabla\phi)^{2}]+V(\phi)\eea
\begin{equation}
F(\phi,\tilde{\phi})=\phi\tilde{\phi}^{2}+\tilde{\phi}^{3}\end{equation}
\begin{equation}
\pi=\tilde{\phi}^{2}\end{equation}
\begin{equation}
-\tilde{\pi}=2\phi\tilde{\phi}+3\tilde{\phi}^{2}\end{equation}
\bea
\tilde{\mathcal{H}}
 & = & \frac{1}{2}[\tilde{\phi}^{4}+(\nabla[\frac{-\tilde{\pi}-3\tilde{\phi}^{2}}{2\tilde{\phi}}])^{2}]+V(\frac{-\tilde{\pi}-3\tilde{\phi}^{2}}{2\tilde{\phi}}) \ \ \ \ \ \ \
 \eea
Hence, we have\begin{equation}
\dot{\tilde{\phi}}=\frac{\delta}{\delta\tilde{\pi}}\int d^{3}x\tilde{\mathcal{H}}=\frac{-1}{2\tilde{\phi}}V'(\frac{-\tilde{\pi}-3\tilde{\phi}^{2}}{2\tilde{\phi}})-\frac{1}{2\tilde{\phi}}\nabla^{2}[\frac{\tilde{\pi}+3\tilde{\phi}^{2}}{2\tilde{\phi}}]
\end{equation}
which allows us to express $\tilde{\pi}$ in terms of $\tilde{\phi}$
and $\dot{\tilde{\phi}}$, but only at the expense of giving up locality
(i.e. one must solve an elliptic PDE). This is the main qualitative difference
between Lagrangian densities obtained from the more general canonical
transformations and local field redefinitions.

\bibliographystyle{apsrev}
\bibliography{bib}

\end{document}